# The Functional Consequences of Variation in Transcription Factor Binding

Darren A. Cusanovich[1], Bryan Pavlovic[1,2], Jonathan K. Pritchard[1,2,3], Yoav Gilad[1]

[1]Department of Human Genetics and [2]Howard Hughes Medical Institute, University of Chicago, Chicago, IL, 60637, USA

[3]Departments of Genetics and Biology and Howard Hughes Medical Institute, Stanford University, Stanford, CA, 94305, USA

Email addresses:

DAC: cusanovich@uchicago.edu; JKP: pritch@stanford.edu; YG: gilad@uchicago.edu

Corresponding authors: DAC, JKP, and YG

**Abstract**

One goal of human genetics is to understand how the information for precise and dynamic gene expression programs is encoded in the genome. The interactions of transcription factors (TFs) with DNA regulatory elements clearly play an important role in determining gene expression outputs, yet the regulatory logic underlying functional transcription factor binding is poorly understood. Many studies have focused on characterizing the genomic locations of TF binding, yet it is unclear to what extent TF binding at any specific locus has functional consequences with respect to gene expression output. To evaluate the context of functional TF binding we knocked down 59 TFs and chromatin modifiers in one HapMap lymphoblastoid cell line. We then identified genes whose expression was affected by the knockdowns. We intersected the gene expression data with transcription factor binding data (based on ChIP-seq and DNase-seq) within 10 kb of the transcription start sites of expressed genes. This combination of data allowed us to infer functional TF binding. On average, 14.7% of genes bound by a factor were differentially expressed following the knockdown of that factor, suggesting that most interactions between TF and chromatin do not result in measurable changes in gene expression levels of putative target genes. We found that functional TF binding is enriched in regulatory elements that harbor a large number of TF binding sites, at sites with predicted higher binding affinity, and at sites that are enriched in genomic regions annotated as "active enhancers".

**Author Summary**

An important question in genomics is to understand how a class of proteins called "transcription factors" controls the expression level of other genes in the genome in a cell-type-specific manner – a process that is essential to human development. One major approach to this problem is to study where these transcription factors bind in the genome, but this does not tell us about the effect of that binding on gene expression levels and it is generally accepted that much of the binding does not strongly influence gene expression. To address this issue, we artificially reduced the concentration of 59 different transcription factors in the cell and then examined which genes were impacted by the reduced transcription factor level. Our results implicate some attributes that might influence what binding is functional, but they also suggest that a simple model of functional vs. non-functional binding may not suffice.

**Introduction**

Understanding the regulatory logic of the genome is critical to understanding human biology. Ultimately, we aim to be able to predict the expression pattern of a gene based on its regulatory sequence alone. However, the regulatory code of the human genome is much more complicated than the triplet code of protein coding sequences, and is highly context-specific, depending on cell-type and other factors [1]. In addition, regulatory regions are not necessarily organized into discrete, easily identifiable regions of the genome and may exert their influence on genes over large genomic distances [2]. Consequently, the rules governing the sequence specificity as well as the functional output of even the most common regulatory interactions, such as interactions between transcription factors and the genome, are not yet fully understood.

To date, genomic studies addressing questions of the regulatory logic of the human genome have largely taken one of two approaches. On the one hand are studies aimed at collecting transcription factor binding maps using techniques such as ChIP-seq and DNase-seq [3–6]. These studies are mainly focused on identifying the specific genomic locations and DNA sequences associated with transcription factor binding and histone modifications. On the other hand are studies aimed at mapping various quantitative trait loci (QTL), such as gene expression levels (eQTLs) [7], DNA methylation (meQTLs) [8] and chromatin accessibility (dsQTLs) [9]. These studies are mainly focused on identifying specific genetic variants that functionally impact gene regulation. Cumulatively, binding map studies and QTL map studies have led to many insights into the principles and mechanisms of gene regulation [7,10–12].

However, there are questions that neither mapping approach on its own is well equipped to address. One outstanding issue is the fraction of factor binding in the genome that is "functional", which we define here to mean that disturbing the protein-DNA interaction leads to a measurable downstream effect on gene regulation. (Note that we do not concern ourselves with the question of whether the regulatory outcome and/or the interaction are evolving under natural selection). An experimental technique that could help address this issue is transcription factor knockdown. In knockdown experiments, the RNA interference pathway is employed to greatly reduce the expression level of a specific target gene by using small interfering RNAs (siRNAs). The cellular or organismal response to the knockdown can then be measured (e.g. [13]). Instead of measuring a cellular phenotype, one can collect RNA after the knockdown and measure global changes in gene expression patterns after specifically attenuating the expression level of a given factor.

Combining a TF knockdown approach with TF binding data can help us to distinguish functional binding from non-functional binding. This approach has previously been applied to the study of human TFs (e.g. [14–16]), although for the most part studies have only focused on the regulatory relationship of a single factor with its downstream targets. The FANTOM consortium previously knocked down 52 different transcription factors in the THP-1 cell line [17], an acute monocytic leukemia-derived cell line, and used a subset of these knockdowns to validate certain regulatory predictions based on binding motif enrichments [18]. However, the amount of transcription factor binding information available for the THP-1 cell line is limited (it is not a part of the ENCODE reference lines).

Many groups, including our own, have previously studied the regulatory architecture of gene expression in the model system of HapMap lymphoblastoid cell lines (LCLs) using both binding map strategies [3,19,20] and QTL mapping strategies [7,9]. As a complement to that work, we sought to use knockdown experiments targeting transcription factors in a HapMap LCL to refine our understanding of the gene regulatory circuitry of the human genome. We integrated the results of the knockdown experiments with previous data on transcription factor binding to better characterize the regulatory targets of 59 different factors and to learn when a disruption in transcription factor binding is most likely to be associated with variation in the expression level of a nearby gene.

**Results**

Our goal was to better characterize gene regulation by transcription factors (TFs). To do so, we measured the impact that knocking down the expression level of TFs and chromatin modifiers had on global gene expression levels in a single HapMap LCL (GM19238). As a first step, we used a high-throughput pipeline to screen siRNAs targeting 112 TFs for their efficiency in knocking down the target transcript (see Table S1 for a list of factors). We evaluated the knockdown efficiency using qPCR to measure transcript levels of the targeted gene in RNA samples extracted 48 hours after the siRNA transfection. Based on the qPCR results, we chose to focus on 59 TFs and chromatin modifiers (see Methods for specific details).

We repeated the knockdown experiment for the 59 factors in triplicate, and collected RNA 72 hours after transfection for gene expression analysis using Illumina HT-12 microarrays. This time point was chosen to provide ample time for the transcript knockdown to impact the protein level of the targeted factor [21,22]. All factors were knocked down in independent cultures of the same LCL. Gene expression levels following the knockdown were compared to expression data collected from six samples that were transfected with negative control siRNA. The expression data from all samples were normalized together using quantile normalization followed by batch correction using the RUV-2 method. We then performed several quality control analyses to confirm that the quality of the data was high, that there were no outlier samples, and that the normalization methods reduced the influence of confounders as much as possible (see Methods, Table S2 and Figures S1-S5). Following these steps, we were able to consider expression data for 7,139 – 8,249 genes (depending on the TF knockdown experiment) that showed detectable expression on all of the knockdown arrays or all of the control arrays (Figure S6). In order to identify genes that were expressed at a significantly different level in the knockdown samples compared to the negative controls, we used likelihood-ratio tests within the framework of a fixed effect linear model (Figure 1, see Methods for details).

**Knockdown effect on global gene expression levels**

Following normalization and quality control of the arrays, we identified genes that were differentially expressed between the three knockdown replicates of each factor and the six controls. Depending on the factor targeted, the knockdowns resulted in between 39 and 3,892 differentially expressed genes at an FDR of 5% (Figure 1B; see Table S3 for a summary of the results). The knockdown efficiency for the 59 factors ranged from 50% to 90% (based on qPCR; Table S1). The qPCR measurements of the knockdown level were significantly correlated with estimates of the TF expression levels based on the microarray data ($P = 0.001$; Figure 1C). Reassuringly, we did not observe a significant correlation between the knockdown efficiency of a given factor and the number of genes classified as differentially expressed following the knockdown experiment (this was true whether we estimated the knockdown effect based on qPCR ($P = 0.10$; Figure 1D) or microarray ($P = 0.99$; not shown) data. Nor did we observe a correlation between variance in qPCR-estimated knockdown efficiency (between replicates) and the number of genes differentially expressed ($P = 0.94$; Figure 1E). We noticed that the large variation in the number of differentially expressed genes extended even to knockdowns of factors from the same gene family. For example, knocking down *IRF4* (with a knockdown efficiency of 86%) resulted in 3,892 differentially expressed genes (including *IRF4*), while knocking down *IRF3* (with a knockdown efficiency of 91%), a paralog of *IRF4* [23], only significantly affected the expression of 113 genes (including *IRF3*).

Because we knocked down 59 different factors in this experiment we were able to assess general patterns associated with the perturbation of transcription factors beyond merely the number of affected target genes. Globally, despite the range in the number of genes we identified as differentially expressed in each knockdown, the effect sizes of the differences in expression were relatively modest and consistent in magnitude across all knockdowns. The median effect size for genes classified as differentially expressed at an FDR of 5% in any knockdown was a 9.2% difference in expression level between the controls and the knockdown (Figure 2), while the median effect size for any individual knockdown experiment ranged between 8.1% and 11.0%.

To further evaluate the biological implications of our observations, we used the Gene Ontology (GO) [24] annotations to identify functional categories enriched among genes that were classified as differentially expressed following the knockdown experiments. In general, the differentially expressed genes tend to be annotated within pathways that fit well with what is already known about the biology of each of the 59 factors (Table S4). For example, differentially expressed genes following the knockdowns of both *IRF4* (3,892 genes differentially expressed) and *IRF9* (243 genes differentially expressed) are enriched for many immune response annotations. However, differentially expressed genes in the *IRF4* knockdown are enriched for both type I and II interferon signaling pathways, among other pathways, consistent with the known role of *IRF4* in immune responses [25]. Genes differentially expressed in the *IRF9* knockdown are enriched for type I interferon responses (among other pathways) but not type II responses, which is again consistent with the known biology [26]. As another example, knocking down *SREBF2* (1,286 genes differentially expressed), a key regulator of cholesterol homeostasis [27], results in changes in the expression of genes that are significantly enriched for cholesterol and sterol biosynthesis annotations. While not all factors exhibited striking enrichments for relevant functional categories and pathways, the overall picture is that perturbations of many of the factors primarily affected pathways consistent with their known biology.

**A combined analysis of factor binding and gene expression data**

In order to assess functional TF binding, we next incorporated binding maps together with the knockdown expression data. In particular, we combined binding data based on DNase-seq footprints in 70 HapMap LCLs, reported by Degner et al. [9] (Table S5) and from ChIP-seq experiments in LCL GM12878, published by ENCODE [3]. We were thus able to obtain genome-wide binding maps for a total of 131 factors that were either directly targeted by an siRNA in our experiment (29 factors) or were differentially expressed in one of the knockdown experiments (see Methods for more details). We classified a gene as a bound target of a particular factor when binding of that factor was inferred within 10kb of the transcription start site (TSS) of the target gene. Using this approach, we found that the 131 TFs were bound in proximity to a median of 1,922 genes per factor (range 11 to 7,053 target genes; Figure S7A; only the 8,872 genes expressed in at least one knockdown experiment were considered for this analysis). Target genes were bound by a median of 34 different factors (range 0 to 96; Figure S7B; only 288 genes expressed in our experiments were not classified as bound targets of any of the 131 factors considered).

We considered binding of a factor to be functional if the target gene was differentially expressed after perturbing the expression level the bound transcription factor. We then asked about the concordance between the transcription factor binding data and the knockdown expression data. Specifically, we studied the extent to which differences in gene expression levels following the knockdowns might be predicted by binding of the transcription factors within the putative regulatory regions of the responsive genes. Likewise, we asked what proportion of putative target (bound) genes of a given TF were also differentially expressed following the knockdown of the factor.

We performed this analysis in two stages. First, we only considered binding data for the specific TF that was knocked down in each experiment (binding data was available for 29 different factors). In general, we found that the number of differentially expressed genes following the knockdowns was positively correlated with the number of bound target genes by these 29 factors (Spearman's $\rho$ = 0.45; permutation $P$ = 0.015). For 12 of the 29 knockdowns, we observed significant overlaps between binding and differential expression (Fisher's exact test; $P$ < 0.05). We also found that between 3.4-75.9% (median = 32.3%) of differentially expressed genes were bound by the TF in a given knockdown (mean relative enrichment = 1.08). Perhaps somewhat less expected, we found that between 46.4% and 99.1% (median = 88.9%) of the binding was apparently not functional, namely it was not associated with changes in gene expression levels. This observation is robust with respect to the size of the window we used to classify genes as bound by a factor (range 1-20kb from the TSS; Table S6). It is also consistent with our previous findings that most DNase-I sensitive QTLs are not also classified as eQTLs [9].

We next considered the expression data in the context of the binding data for both the knocked down TFs and any other TF whose expression level was indirectly affected by the knockdown. We again examined the overlap between binding and differential expression (Figure 3). Considering only the expressed genes in each experiment, the fraction of genes bound (by any TF whose expression has changed) ranged between 16.2% and 95.3% with a median of 85.4% (Figure 3B). However, the fraction of bound genes that were also differentially expressed in a given experiment was generally quite low (median = 7.9%; Figure 3C; mean relative enrichment = 1.02), with significant ($P$ < 0.05) overlap between bound and differentially expressed genes seen at only 13 of the 59 knockdown experiments. Even if we relaxed the statistical threshold with which we classify genes as differentially expressed four-fold (to an FDR of 20%), a majority of bound genes still failed to show significant evidence of differential expression (median = 68.8%; Figure 3C; mean relative enrichment = 1.01). The discrepancy in the number of genes bound by a particular factor (or the TFs it regulates) and the number of differentially expressed genes in a knockdown experiment begged the question of whether any characteristics of factor binding might distinguish functionally bound target genes. In order to address this question, we examined a variety of features.

**Functional factor binding is enriched in enhancer chromatin states**

First, focusing only on the binding sites classified using the DNase-seq data (which were assigned to a specific instance of the binding motif, unlike the ChIP data), we examined sequence features that might distinguish functional binding. In particular, we considered whether binding at conserved sites was more likely to be functional (estimating conservation by using PhastCons 46 way placental scores [28]) and we also considered whether binding sites that better matched the known PWM for the factor were more likely to be functional. Interestingly, we did not observe a significant shift in the conservation of functional binding sites (Wilcoxon rank sum $P$ = 0.34), but we did observe that binding around differentially expressed genes occurred at sites that were significantly better matches to the canonical binding motif ($P < 10^{-8}$), although the absolute difference in PWM score was very small.

Next, considering bound targets determined from either the ChIP-seq or DNase-seq data, we observed that differentially expressed genes were associated with both a higher number of binding events for the relevant factors within 10kb of the TSS ($P < 10^{-16}$; Figure 4A) as well as with a larger number of different binding factors (considering the siRNA-targeted factor and any TFs that were DE in the knockdown; $P < 10^{-16}$; Figure 4B). We hoped to distinguish between coordinated co-regulation of the factors and generally higher levels of binding nearby differentially expressed genes. To do so, we asked whether the genes differentially expressed in common between any two knockdown experiments were more likely to be co-occupied by the same transcription factors (considering only transcription factors whose expression was affected by the knockdown). Binning all pairwise comparisons between knockdown experiments based on the fraction of differentially expressed transcription factors in common, we observed that enrichment for functional co-occupancy increased proportionally to the fraction of TFs in common (Figure 4C). This suggests that co-regulation is at least partially responsible for the increased numbers of factors binding near differentially expressed genes.

We proceeded by examining the distribution of binding about the TSS. Most factor binding was concentrated near the TSS whether or not the genes were classified as differentially expressed (Figure 5A). However, surprisingly, the distance from the TSS to the binding sites was significantly longer for differentially expressed genes ($P < 10^{-16}$; Fig 5B). We then investigated the distribution of factor binding across various chromatin states, as defined by Ernst et al. [11]. This dataset lists regions of the genome that have been assigned to different activity states based on ChIP-seq data for various histone modifications and *CTCF* binding. For each knockdown, we separated binding events by the genomic state in which they occurred and then tested whether binding in that state was enriched around differentially expressed genes. After correcting for multiple testing, 19 knockdowns showed significant enrichment for binding in "strong enhancers" around genes that were differentially expressed and four knockdowns had significant enrichments for "weak enhancers". Further, eight knockdowns showed significant depletion of binding in "active promoters" of genes that were differentially expressed and six knockdowns had significant depletions for "transcription elongation".

**The direction of expression change**

Finally, we asked whether the factors tended to have a consistent effect (either up- or down-regulation) on the expression levels of genes they purportedly regulated. Perhaps surprisingly, all factors we tested are associated with both up- and down-regulation of downstream targets (Figure 6). A slight majority of downstream target genes were expressed at higher levels following the knockdown for 15 of the 29 factors for which we had binding information (Figure 6B). The factor that is associated with the largest fraction (68.8%) of up-regulated target genes following the knockdown is *EZH2*, the enzymatic component of the Polycomb group complex. On the other end of the spectrum was *JUND*, a member of the AP-1 complex, for which 66.7% of differentially expressed targets were down-regulated following the knockdown. The remaining 27 factors (with a median of 170 direct targets) all show a more even balance between up- and down-regulated targets. These trends are consistent when we considered all genes that were differentially expressed following a knockdown (not just the genes that were also bound by the knocked down factor). We observed that an average of 51.9% of downstream differentially expressed genes had elevated expression levels following knockdown of the transcription factors. Furthermore, for 39 of the experiments a slight majority of differentially expressed genes following knockdown of the factor were up-regulated, while in only 17 of the experiments were the majority of differentially expressed genes down-regulated following the knockdown.

**Discussion**

The question of the distinction between functional and non-functional DNA in the human genome has attracted considerable attention recently. Some have argued that any base that comes into contact with a protein or RNA molecule at some point in its lifespan should be considered functional [3], which is an extremely inclusive definition. Conversely, others have argued that only the bases of the genome that evolve under natural selection pressure and can be clearly ascribed to a particular phenotype should be considered functional [29]. We agree that functional interpretation makes more sense in an evolutionary context, but we also recognize that our ability to identify the signatures of natural selection (especially positive selection) is limited. Instead, we propose an experimentally tractable definition of functional interactions between transcription factors and DNA, which allowed us to make several novel observations regarding the nuanced regulatory logic of the human genome.

**Characterizing functional transcription factor binding**

Nearly all expressed genes in the LCL we worked with are bound within 10 kb of their TSS by at least one of the 131 TFs for which we were able to obtain binding data. Yet, the regulation of the vast majority of target genes is not affected by perturbations to the expression levels of the TFs. Our observations suggest that it may ultimately be possible to predict functional transcription factor binding based on the biological context, yet since bound genes are only modestly enriched among those that are differentially expressed, an effective classifier may be difficult to develop. In that context, several of the associations we observed might seem counterintuitive from a purely biochemical perspective but are consistent with our definition of functional TF binding as participating in gene regulation. In particular, from a biochemical perspective, binding at stronger motifs might be expected to be less affected by a decrease in factor concentration in the cell (following the knockdown) and regulatory regions with more binding sites and a larger number of bound factors might be expected to be less influenced by the perturbation of one single factor. Yet, we observed the opposite patterns: Functional binding is associated with stronger binding motifs and greater levels of factor binding near differentially expressed genes. Viewed from an evolutionary rather than purely biochemical perspective, these observations are quite in tune with expectations. In other words, genomic regions of functional importance evolve to ensure factor binding. This can be accomplished by selection for different properties, including increased affinity of the binding site and more cooperative binding.

Our results also indicate that binding in the context of certain chromatin states was more likely to be functional. For 19 of the 56 factors that we knocked down and for which we were able to obtain binding data (on either direct or indirect bound targets), there was a significant enrichment of binding in "strong enhancers" near differentially expressed genes, which is consistent with our observations that functional binding occurs further from the TSS (namely, not in promoter regions). While further experiments are required before we can put forward a more concrete explanation, these observations suggest that binding at active promoters may be buffered against acute changes in transcription factor concentration. This may also explain why most of the effect sizes associated with differences in gene expression levels following the knockdowns were relatively modest. While there is compelling evidence for our inferences, the current chromatin functional annotations do not fully explain the regulatory effects of the knockdown experiments. For example, the enrichments for binding in "strong enhancer" regions of the genome range from 7.2% to 50.1% (median = 19.2%), much beyond what is expected by chance alone, but far from accounting for all functional binding.

**Evaluating the impact of transcription factor binding on direct targets**

In addition to considering the distinguishing characteristics of functional binding, we also examined the direction of effect that perturbing a transcription factor had on the expression level of its direct targets. We specifically addressed whether knocking down a particular factor tended to drive expression of its putatively direct (namely, bound) targets up or down, which can be used to infer that the factor represses or activates the target, respectively. Transcription factors have traditionally been thought of primarily as activators [11,30,31], and previous work from our group is consistent with that notion [9]. Surprisingly, the most straightforward inference from the present study is that many of the factors function as repressors at least as often as they function as activators. For example, we inferred that *EZH2* had a negative regulatory relationship with the largest fraction of direct targets (68.8%), while *JUND* seemed to have a positive regulatory relationship with the largest fraction of direct targets (66.7%). These particular observations seem consistent with the known role of *EZH2* as the active member of the Polycomb group complex PC2 [32] and the biochemical characterization of the AP-1 complex (of which *JUND* is a component) as a transactivator [33]. More generally, however, our results, combined with the previous work from our group and others (e.g. [9,11]) make for a complicated view of the role of transcription factors in gene regulation as it seems difficult to reconcile the inference from previous work that many transcription factors should primarily act as activators with the results presented here.

One somewhat complicated hypothesis, which nevertheless can resolve the apparent discrepancy, is that the "repressive" effects we observe for known activators may be at sites in which the activator is acting as a weak enhancer of transcription and that reducing the cellular concentration of the factor releases the regulatory region to binding by an alternative, stronger activator. We believe that this may alleviate the apparently contradictory interpretations of transcription factor activity, although additional work on this topic is needed.

**Possible caveats and future studies**

There are important caveats that should be kept in mind when interpreting our results. Care must be taken in interpreting the lack of evidence for gene expression differences in our system as previous studies have suggested that the specific effects of knockdowns may be difficult to detect because of redundancy in gene regulatory mechanisms [34,35]. In other words, we may be underestimating the fraction of bound genes that are functionally regulated. We have taken measures to control for this as much as possible. First, integrating our results with binding data allows us to focus our analysis on the most likely direct targets. Second, the consistency of our results across a variety of factors (including factors with very different numbers of differentially expressed genes following the knockdown), suggests that technical explanations for our observations are unlikely.

In light of our observations a reassessment of our estimates of binding may be warranted. In particular, because functional binding is skewed away from promoters (our system is apparently not well-suited to observe functional promoter binding, perhaps because of protection by large protein complexes), a more conservative estimate of the fraction of binding that is indeed functional would not consider data within the promoter. Importantly, excluding the putative promoter region from our analysis (i.e. only considering a window >1kb from the TSS and <10kb from the TSS) does not change our conclusions. Considering this smaller window, a median of 67.0% of expressed genes are still classified as bound by either the knocked down transcription factor or a downstream factors that is differentially expressed in each experiment, yet a median of only 8.1% of the bound genes are also differentially expressed after the knockdowns.

Further work in this field is clearly justified. Much of what distinguishes functional binding (as we define it) has yet to be explained. Furthermore, we are unable to explain much of the differential expression observed in our experiments by the presence of least one relevant binding event. To address these issues, more factors should be perturbed to further evaluate the robustness of our results and to add insight. Together, such studies will help us develop a more sophisticated understanding of functional transcription factor binding in particular, the gene regulatory logic more generally.

**Methods**

**Cell Culture and siRNA Transfections**

The cell line (GM19238) was cultured at 5% $CO_2$ and 37°C in RPMI 1640 medium supplemented with 2mM L-glutamine and 15% fetal bovine serum, per Coriell's recommendations (http://ccr.coriell.org/Sections/Support/Global/Lymphoblastoid.aspx?PgId=213). The medium was also supplemented with 100 IU/ml penicillin and 100 μg/ml streptomycin. Cells were counted and split three times a week to 350,000 cells/ml. Transfections were performed with the Lonza 96-well nucleofector system, using transfection solution SF and transfection program DN-100. On-Target SmartPool siRNAs from Dharmacon were used to knockdown target genes. For each transfection, one million cells were transfected with 50 pmol siRNA. After transfection, cells were plated in 1 ml of medium in 96-well plates and incubated for 48. At the 48 hour time point, 500 μl of cell culture were removed for RNA extraction (for qPCR), 100 μl of fresh medium was added to each well and plates were incubated for an additional 24 hours. At the 72 hour time point, the remaining culture was pelleted for RNA extraction (for array hybridization). RNA from all timepoints was extracted using the RNeay Plus 96 kit (Qiagen). The control siRNAs consisted of a pool of four siRNAs specifically designed not to target any human or mouse gene. In addition to two replicates of the negative control pool, we also transfected each of the negative control siRNAs independently for a total of six negative control samples. We learned from an initial pilot experiment that there were strong batch effects between rounds of transfection and so we included negative control transfection in parallel with all three batches of transfection.

The transfections were conducted in two phases. For the first phase, we screened siRNAs for their knockdown efficiency under our experimental conditions by transfecting cells and extracting RNA 48 hours later. qPCR was performed with SYBR Green and custom primers (Table S1). Knockdown efficiency was assessed relative to a sample transfected with the negative control siRNA pool. We used a relative quantification approach, referencing the *POLR2C* control gene and using the DART-PCR method [36] to determine PCR efficiency. For the second phase, those siRNA transfections resulting in ≥50% knockdown (arbitrary cutoff) were transfected again and RNA was extracted at both 48 hours (to confirm the knockdown level by qPCR) and 72 hours (for hybridization to microarrays). Two genes (*NFYC* and *ZHX2*) were not knocked down by 50% in the screen but were included in the final transfection where they did reach 50% (in all three replicates) and were therefore included in the microarray experiment. Additionally, 16 factors that passed the screen were not knocked down by 50% at 48 hours in all three replicates in the final transfection and were not included in the microarray experiment (Table S1). The quality and concentration of all RNA samples were measured using the Agilent 2100 Bioanalyzer.

**Transcription Factors and Chromatin Modifiers Targeted by siRNA**

*ARNTL2*, *BATF*, *BCL3*, *CEBPG*, *CEBPZ*, *CLOCK*, *CREBBP*, *DIP2B*, *E2F1*, *E2F4*, *E2F6*, *EP300*, *ESRRA*, *EZH2*, *FOXA3*, *GTF2B*, *HCST*, *HOXB7*, *IKZF3*, *IRF3*, *IRF4*, *IRF5*, *IRF7*, *IRF8*, *IRF9*, *JUND*, *KLF13*, *LCORL*, *NFE2L1*, *NFKB2*, *NFX1*, *NFYC*, *NR1D2*, *NR2F6*, *NR3C1*, *PAX5*, *POU2F1*, *POU2F2*, *RAD21*, *RDBP*, *RELA*, *RELB*, *RXRA*, *SKIL*, *SP1*, *SP3*, *SREBF2*, *STAT2*, *STAT6*, *TAF1*, *TCF12*, *TFDP1*, *TFDP2*, *TFE3*, *USF1*, *WHSC1*, *YY1*, *ZBTB38*, *ZHX2*.

**Gene Expression Arrays and Normalization**

Samples were hybridized to Illumina HT-12 v4R2 arrays in two batches. For the first batch, 69 samples (150 ng of total RNA), representing 21 knockdowns and one set of controls, were sent to the UCLA Southern California Genotyping Consortium where the RNA was converted to cRNA and then hybridized to arrays using the standard protocol. For the second batch, 132 samples (1 µg of total RNA), representing 40 knockdowns and two sets of controls, were sent to the University of Chicago Functional Genomics Core where the RNA was converted to cRNA and then hybridized to the arrays using the standard protocol. Both cores returned raw probe intensities to us. The batch aspect of the study design is clearly less than optimal, but it was a result of necessary practical considerations. As mentioned above, we included a full set of negative controls with each batch. We also accounted for batch effects explicitly in our analysis (see below).

Before processing the arrays, we determined the set of usable probes. To do this, we first mapped probes to the hg19 reference genome using the BWA alignment program [37] and only retained probes that mapped perfectly to the genome (probes spanning introns were discarded). We then removed probes that mapped perfectly to a single site in the genome and also mapped to a second site allowing for one mismatch. As GM19238 was derived from a female donor, we excluded all probes for Y chromosome genes. We then removed probes that contained a SNP for which GM19238 was heterozygous based on 1000 Genomes data [38] (http://ftp.1000genomes.ebi.ac.uk/vol1/ftp/pilot_data/paper_data_sets/a_map_of_human_variation/trio/snps/) to avoid detecting spurious results from an interaction between allele specific expression and probe effects influencing hybridization intensities. After this series of exclusions, for each gene that was assayed by more than one remaining probe, we chose the 3'-most probe to represent the gene. We then excluded probes for genes that were not expressed (detection $P < 0.01$) in the knockdown triplicates or the full set of controls (18 arrays) for each experiment.

We log-transformed and quantile normalized the data from all arrays together using the 'lumi' Bioconductor package [39,40] in the R statistical environment. Our initial analysis indicated that this was not sufficient to correct for all of the batch effects present between different rounds of transfection (Figure S5), so we also used the RUV-2 method [41] to further adjust the data. To do so, we defined a set of genes that should not be differentially expressed under any conditions in our experiment as the basis for the RUV-2 correction.

While the choice of unaffected genes might not be straightforward in all experimental settings, the fact that we had 59 different knockdown experiments (and controls) across three rounds of transfections provided an opportunity to define a reasonable set of control probes. Specifically, we created a list of the 2,000 least variable probes (considering only probes for genes that passed our expression threshold) for each of the batches of arrays separately. We then intersected the three lists to obtain a high confidence list of 787 invariant genes. These are the genes with the least evidence of difference in expression levels in any knockdown experiment.

We also needed to specify the number of components to regress out. Again, this choice may not always be straightforward, however, we had two knockdown experiments (*IRF5* and *SP1*) that were each repeated across different batches. We thus maximized the correlation between these experiments in deciding on the number of components to remove. We found that removing eight components resulted in zero genes identified as differentially expressed (at an FDR of 5%) between replicate experiments (or the control arrays by themselves, across batches) and had the maximum number of differentially expressed genes replicated between the two experiments for both *IRF5* and *SP1* (Figure S2; these models were fit with the 'limma' package [42]). Lastly, we averaged each of the negative control knockdowns across each of the time points in order to have a single set of controls with which to compare each of the knockdowns. In addition to examining heatmaps for the arrays (Figures S1, S3), the quality of each array was assessed by relative log expression ('RLE') plots (Figure S4; [43,44]) and PCA (Figure S5, Table S2). For the remainder of the analyses, we only used the *IRF5* and *SP1* experiment that had the greater knockdown efficiency.

We note that the top principle components of the adjusted data are still correlated with the microarray chip (Table S2), but we decided that further correction was not helpful for several reasons. First, the amount of variance captured by any one component is quite low after RUV-2 adjustment (Table S2). Second, our analysis indicated that removing additional components with RUV-2 reduced the correlations between replicate knockdown experiments (Figure S2). Third, we randomized samples across chips so as to avoid the introduction of a bias in our results based on such an effect.

**Determining Differentially Expressed Genes**

In order to identify differentially expressed genes in each knockdown, we compared the expression level of each gene on the three knockdown arrays to its expression level on the six control arrays. We used a likelihood-ratio test within the framework of a fixed-effect linear model:

$$Y_{ij} = \mu_i + \beta_j X_i + \varepsilon_{ij}$$

Here, $Y_{ij}$ is the expression level of gene i on array j. $\mu_i$ is the mean expression level for gene i. $\beta_j$ is the status of array j (either "knockdown" or "control") and $X_i$ is the knockdown effect for gene i. $\varepsilon_{ij}$ is the error term for gene i on array j. For each gene, we compared a model with a $\beta_j X_i$ term to the nested model with no $\beta_j X_i$ term using a likelihood ratio test to determine which model fit the data better. To adjust for multiple testing within each knockdown experiment, we calculated the q-value for each gene using the Storey and Tibshirani method [45] as provided in the 'qvalues' R package.

**Gene Ontology Analysis**

For each knockdown, we also assessed whether there were any Gene Ontology categories (either from the "MF" or "BP" categories) that were overrepresented among the differentially expressed genes. To do so, we used the 'topGO' package and the 'org.Hs.eg' database in R. For each knockdown, we tested both "MF" and "BP" categories, combined the results, and filtered out any categories without a single gene differentially expressed. P-values were adjusted for multiple testing using the 'BH' method in the 'p.adjust' function in R.

**Combining with other genomic datasets**

After identifying differentially expressed genes, we intersected the gene expression data with factor binding data from both DNase-seq experiments and ChIP-seq experiments. We considered binding data within a fixed window around the TSS of each gene. To determine a single TSS for each gene, we used the ENCODE CAGE data downloaded from UCSC [3,46]. For each gene, we chose the TSS with the highest CAGE score as the reference TSS for that gene unless there was a tie between multiple TSSs, in which case we used the midpoint between these TSSs as the reference TSS. The DNase data was from a previous study conducted in our lab [9]. Binding was determined using the Centipede algorithm [4] on DNase-seq data from 70 Yoruba HapMap cell lines. For each factor expressed in our experiments, we classified all sites with a Centipede posterior probability greater than 0.95 as bound. The binding was originally mapped to the hg18 reference genome, so we used liftOver (http://hgdownload.cse.ucsc.edu/admin/exe/) to convert the coordinates to h19.

For the ChIP data, we downloaded all ENCODE ChIP-seq data for GM12878 called using the SPP peak caller, except for the *POL2* datasets and an *NFKB* dataset that was collected following tumor necrosis factor-α stimulation [3] (http://ftp.ebi.ac.uk/pub/databases/ensembl/encode/integration_data_jan2011/byDataType/peaks/jan2011/spp/optimal/hub/). For each gene that we were able to link to a ChIP dataset or a TRANSFAC binding motif [47], we combined all binding records and then considered the union of that set as the binding profile for the factor. After obtaining the union set, we calculated the midpoint for each discrete binding record and used the midpoints as the estimated binding location in all subsequent analyses.

Using this approach, we obtained binding data for 201 factors; 138 of these factors were represented by a usable probe on the array and were expressed in at least one of the knockdown experiments. 131 of the factors were differentially expressed in at least one knockdown experiment. For DNase-based binding sites, we also evaluated the PhastCons alignment score [28]. PhastCons 46 way placental wig files were downloaded from UCSC and the average score for each DNase-seq binding site was calculated. For all of the comparisons between functional and non-functional binding, we used the Wilcoxon rank sum test to assess differences.

We also downloaded the Ernst chromatin states [11] from UCSC. This file contained 15 different chromatin states, including two separate categories each for "strong enhancer", "weak enhancer", and "repetitive". We combined each of the replicate states into a single category so that we ended up with 12 distinct chromatin states. To identify which state the binding occurred in, we intersected the binding record midpoints with the chromatin states. We calculated the enrichment or depletion of functional binding in specific chromatin states using a Fisher's Exact Test. All analyses were performed with a combination of BedTools [48,49] and BEDOPS [50] commands, along with custom Python and R scripts.

**Data Availability**

The knockdown data have been deposited in NCBI's Gene Expression Omnibus ([51]; accessible through GEO Series accession number GSE50588 at http://www.ncbi.nlm.nih.gov/geo/query/acc.cgi?acc=GSE50588)

**Financial Disclosure**

Funded by NIH grant HG006123 to YG and by Howard Hughes Medical Institute funds to JKP. DAC is partially supported by NIH grant T32 GM007197. The funders had no role in study design, data collection and analysis, decision to publish, or preparation of the manuscript.

**Acknowledgements**

We thank G. McVicker, R. Pique-Regi, N. Banovich, J. Blischak, J. Roux and S. Lindstrom for helpful discussions about experimental design and analysis and comments on the manuscript, as well as other members of the Gilad and Pritchard labs for assistance throughout. We thank J. DeYoung at the Southern California Genotyping Consortium and P. Faber at the University of Chicago Functional Genomics Core for their outstanding service. We acknowledge the ENCODE consortium for providing access to their data.

# References

1. Jolma A, Yan J, Whitington T, Toivonen J, Nitta KR, et al. (2013) DNA-binding specificities of human transcription factors. Cell 152: 327–339. doi:10.1016/j.cell.2012.12.009.

2. Nobrega MA, Ovcharenko I, Afzal V, Rubin EM (2003) Scanning human gene deserts for long-range enhancers. Science 302: 413. doi:10.1126/science.1088328.

3. Bernstein BE, Birney E, Dunham I, Green ED, Gunter C, et al. (2012) An integrated encyclopedia of DNA elements in the human genome. Nature 489: 57–74. doi:10.1038/nature11247.

4. Pique-Regi R, Degner JF, Pai AA, Gaffney DJ, Gilad Y, et al. (2011) Accurate inference of transcription factor binding from DNA sequence and chromatin accessibility data. Genome Res 21: 447–455. doi:10.1101/gr.112623.110.

5. Song L, Crawford GE (2010) DNase-seq: a high-resolution technique for mapping active gene regulatory elements across the genome from mammalian cells. Cold Spring Harb Protoc 2010: pdb.prot5384. doi:10.1101/pdb.prot5384.

6. Yan J, Enge M, Whitington T, Dave K, Liu J, et al. (2013) Transcription factor binding in human cells occurs in dense clusters formed around cohesin anchor sites. Cell 154: 801–813. doi:10.1016/j.cell.2013.07.034.

7. Gaffney DJ, Veyrieras J-B, Degner JF, Pique-Regi R, Pai AA, et al. (2012) Dissecting the regulatory architecture of gene expression QTLs. Genome Biol 13: R7. doi:10.1186/gb-2012-13-1-r7.

8. Bell JT, Pai AA, Pickrell JK, Gaffney DJ, Pique-Regi R, et al. (2011) DNA methylation patterns associate with genetic and gene expression variation in HapMap cell lines. Genome Biol 12: R10. doi:10.1186/gb-2011-12-1-r10.

9. Degner JF, Pai AA, Pique-Regi R, Veyrieras J-B, Gaffney DJ, et al. (2012) DNase I sensitivity QTLs are a major determinant of human expression variation. Nature 482: 390–394. doi:10.1038/nature10808.

10. Spivakov M, Akhtar J, Kheradpour P, Beal K, Girardot C, et al. (2012) Analysis of variation at transcription factor binding sites in Drosophila and humans. Genome Biol 13: R49. doi:10.1186/gb-2012-13-9-r49.

11. Ernst J, Kheradpour P, Mikkelsen TS, Shoresh N, Ward LD, et al. (2011) Mapping and analysis of chromatin state dynamics in nine human cell types. Nature 473: 43–49. doi:10.1038/nature09906.

12. Gilad Y, Rifkin SA, Pritchard JK (2008) Revealing the architecture of gene regulation: the promise of eQTL studies. Trends Genet 24: 408–415. doi:10.1016/j.tig.2008.06.001.

13. Chia N-Y, Chan Y-S, Feng B, Lu X, Orlov YL, et al. (2010) A genome-wide RNAi screen reveals determinants of human embryonic stem cell identity. Nature 468: 316–320. doi:10.1038/nature09531.

14. Yang A, Zhu Z, Kapranov P, McKeon F, Church GM, et al. (2006) Relationships between p63 binding, DNA sequence, transcription activity, and biological function in human cells. Mol Cell 24: 593–602. doi:10.1016/j.molcel.2006.10.018.

15. Krig SR, Jin VX, Bieda MC, O'Geen H, Yaswen P, et al. (2007) Identification of genes directly regulated by the oncogene ZNF217 using chromatin immunoprecipitation (ChIP)-chip assays. J Biol Chem 282: 9703–9712. doi:10.1074/jbc.M611752200.

16. Xu X, Bieda M, Jin VX, Rabinovich A, Oberley MJ, et al. (2007) A comprehensive ChIP-chip analysis of E2F1, E2F4, and E2F6 in normal and tumor cells reveals interchangeable roles of E2F family members. Genome Res 17: 1550–1561. doi:10.1101/gr.6783507.

17. Kawaji H, Severin J, Lizio M, Waterhouse A, Katayama S, et al. (2009) The FANTOM web resource: from mammalian transcriptional landscape to its dynamic regulation. Genome Biol 10: R40. doi:10.1186/gb-2009-10-4-r40.

18. Suzuki H, Forrest ARR, van Nimwegen E, Daub CO, Balwierz PJ, et al. (2009) The transcriptional network that controls growth arrest and differentiation in a human myeloid leukemia cell line. Nat Genet 41: 553–562. doi:10.1038/ng.375.

19. Cheng C, Alexander R, Min R, Leng J, Yip KY, et al. (2012) Understanding transcriptional regulation by integrative analysis of transcription factor binding data. Genome Res 22: 1658–1667. doi:10.1101/gr.136838.111.

20. Gerstein MB, Kundaje A, Hariharan M, Landt SG, Yan K-K, et al. (2012) Architecture of the human regulatory network derived from ENCODE data. Nature 489: 91–100. doi:10.1038/nature11245.

21. Alemán LM, Doench J, Sharp PA (2007) Comparison of siRNA-induced off-target RNA and protein effects. RNA 13: 385–395. doi:10.1261/rna.352507.

22. De Candia P, Blekhman R, Chabot AE, Oshlack A, Gilad Y (2008) A combination of genomic approaches reveals the role of FOXO1a in regulating an oxidative stress response pathway. PLoS One 3: e1670. doi:10.1371/journal.pone.0001670.

23. Tamura T, Yanai H, Savitsky D, Taniguchi T (2008) The IRF family transcription factors in immunity and oncogenesis. Annu Rev Immunol 26: 535–584. doi:10.1146/annurev.immunol.26.021607.090400.

24. Ashburner M, Ball CA, Blake JA, Botstein D, Butler H, et al. (2000) Gene ontology: tool for the unification of biology. The Gene Ontology Consortium. Nat Genet 25: 25–29. doi:10.1038/75556.

25. Kanno Y, Levi B-Z, Tamura T, Ozato K (2005) Immune cell-specific amplification of interferon signaling by the IRF-4/8-PU.1 complex. J Interferon Cytokine Res 25: 770–779. doi:10.1089/jir.2005.25.770.

26. Tsuno T, Mejido J, Zhao T, Schmeisser H, Morrow A, et al. (2009) IRF9 is a key factor for eliciting the antiproliferative activity of IFN-alpha. J Immunother 32: 803–816. doi:10.1097/CJI.0b013e3181ad4092.

27. Eberlé D, Hegarty B, Bossard P, Ferré P, Foufelle F (2004) SREBP transcription factors: master regulators of lipid homeostasis. Biochimie 86: 839–848. doi:10.1016/j.biochi.2004.09.018.

28. Siepel A, Bejerano G, Pedersen JS, Hinrichs AS, Hou M, et al. (2005) Evolutionarily conserved elements in vertebrate, insect, worm, and yeast genomes. Genome Res 15: 1034–1050. doi:10.1101/gr.3715005.

29. Graur D, Zheng Y, Price N, Azevedo RBR, Zufall RA, et al. (2013) On the immortality of television sets: "function" in the human genome according to the evolution-free gospel of ENCODE. Genome Biol Evol 5: 578–590. doi:10.1093/gbe/evt028.

30. Latchman DS (2001) Transcription factors: bound to activate or repress. Trends Biochem Sci 26: 211–213. doi:10.1016/S0968-0004(01)01812-6.

31. Boyle P, Després C (2010) Dual-function transcription factors and their entourage: unique and unifying themes governing two pathogenesis-related genes. Plant Signal Behav 5: 629–634.

32. Hobert O, Jallal B, Ullrich A (1996) Interaction of Vav with ENX-1, a putative transcriptional regulator of homeobox gene expression. Mol Cell Biol 16: 3066–3073.

33. Hirai SI, Ryseck RP, Mechta F, Bravo R, Yaniv M (1989) Characterization of junD: a new member of the jun proto-oncogene family. EMBO J 8: 1433–1439.

34. Farnham PJ (2009) Insights from genomic profiling of transcription factors. Nat Rev Genet 10: 605–616. doi:10.1038/nrg2636.

35. Biggin MD (2011) Animal transcription networks as highly connected, quantitative continua. Dev Cell 21: 611–626. doi:10.1016/j.devcel.2011.09.008.

36. Peirson SN (2003) Experimental validation of novel and conventional approaches to quantitative real-time PCR data analysis. Nucleic Acids Res 31: 73e–73. doi:10.1093/nar/gng073.

37. Li H, Durbin R (2009) Fast and accurate short read alignment with Burrows-Wheeler transform. Bioinformatics 25: 1754–1760. doi:10.1093/bioinformatics/btp324.

38. Abecasis GR, Altshuler D, Auton A, Brooks LD, Durbin RM, et al. (2010) A map of human genome variation from population-scale sequencing. Nature 467: 1061–1073. doi:10.1038/nature09534.

39. Du P, Kibbe WA, Lin SM (2008) lumi: a pipeline for processing Illumina microarray. Bioinformatics 24: 1547–1548. doi:10.1093/bioinformatics/btn224.

40. Gentleman RC, Carey VJ, Bates DM, Bolstad B, Dettling M, et al. (2004) Bioconductor: open software development for computational biology and bioinformatics. Genome Biol 5: R80. doi:10.1186/gb-2004-5-10-r80.

41. Gagnon-Bartsch JA, Speed TP (2012) Using control genes to correct for unwanted variation in microarray data. Biostatistics 13: 539–552. doi:10.1093/biostatistics/kxr034.

42. Smyth GK (2004) Linear models and empirical bayes methods for assessing differential expression in microarray experiments. Stat Appl Genet Mol Biol 3: Article3. doi:10.2202/1544-6115.1027.

43. Bolstad B, Irizarry R, Åstrand M, Speed T (2003) A comparison of normalization methods for high density oligonucleotide array data based on variance and bias. Bioinformatics 19: 185–193.

44. Brettschneider J, Collin F, Bolstad BM, Speed TP (2008) Quality assessment for short oligonucleotide microarray data. Technometrics 50: 241–264. doi:10.1198/004017008000000334.

45. Storey JD, Tibshirani R (2003) Statistical significance for genomewide studies. Proc Natl Acad Sci 100: 9440–9445. doi:10.1073/pnas.1530509100.

46. Djebali S, Davis CA, Merkel A, Dobin A, Lassmann T, et al. (2012) Landscape of transcription in human cells. Nature 489: 101–108. doi:10.1038/nature11233.

47. Matys V, Kel-Margoulis O V, Fricke E, Liebich I, Land S, et al. (2006) TRANSFAC and its module TRANSCompel: transcriptional gene regulation in eukaryotes. Nucleic Acids Res 34: D108–10. doi:10.1093/nar/gkj143.

48. Quinlan AR, Hall IM (2010) BEDTools: a flexible suite of utilities for comparing genomic features. Bioinformatics 26: 841–842. doi:10.1093/bioinformatics/btq033.

49. Dale RK, Pedersen BS, Quinlan AR (2011) Pybedtools: a flexible Python library for manipulating genomic datasets and annotations. Bioinformatics 27: 3423–3424. doi:10.1093/bioinformatics/btr539.

50. Neph S, Kuehn MS, Reynolds AP, Haugen E, Thurman RE, et al. (2012) BEDOPS: high-performance genomic feature operations. Bioinformatics 28: 1919–1920. doi:10.1093/bioinformatics/bts277.

51. Edgar R (2002) Gene Expression Omnibus: NCBI gene expression and hybridization array data repository. Nucleic Acids Res 30: 207–210. doi:10.1093/nar/30.1.207.

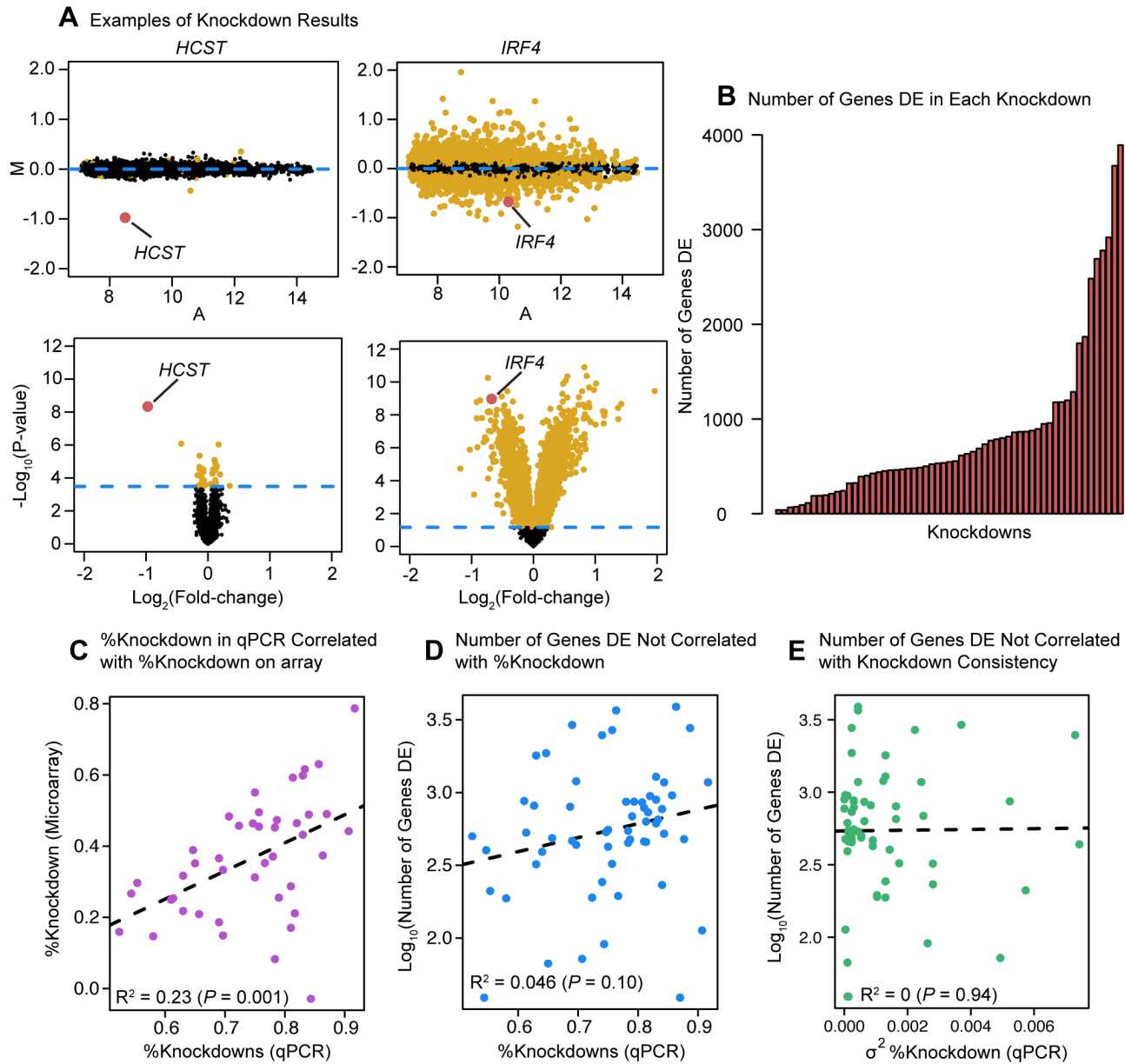

**Figure 1**

**Differential expression analysis.** (a) Examples of differential expression analysis results for the genes *HCST* and *IRF4*. The top two panels are 'MA plots' of the mean $\text{Log}_2$(expression level) between the knockdown arrays and the controls for each gene (x-axis) to the $\text{Log}_2$(Fold-Change) between the knockdowns and controls (y-axis). Differentially expressed genes at an FDR of 5% are plotted in yellow (points 50% larger). The gene targeted by the siRNA is highlighted in red. The bottom two panels are 'volcano plots' of the $\text{Log}_2$(Fold-Change) between the knockdowns and controls (x-axis) to the P-value for differential expression (y-axis). The dashed line marks the 5% FDR threshold. Differentially expressed genes at an FDR of 5% are plotted in yellow (points 50% larger).

The red dot marks the gene targeted by the siRNA. (b) Barplot of number of differentially expressed genes in each knockdown experiment. (c) Comparison of the knockdown level measured by qPCR (RNA sample collected 48 hours post-transfection) and the knockdown level measured by microarray. (d) Comparison of the level of knockdown of the transcription factor at 48hrs (evaluated by qPCR; x-axis) and the number of genes differentially expressed in the knockdown experiment (y-axis). (e) Comparison of the variance in knockdown efficiency between replicates for each transcription factor (evaluated by qPCR; x-axis) and the number of differentially expressed genes in the knockdown experiment (y-axis).

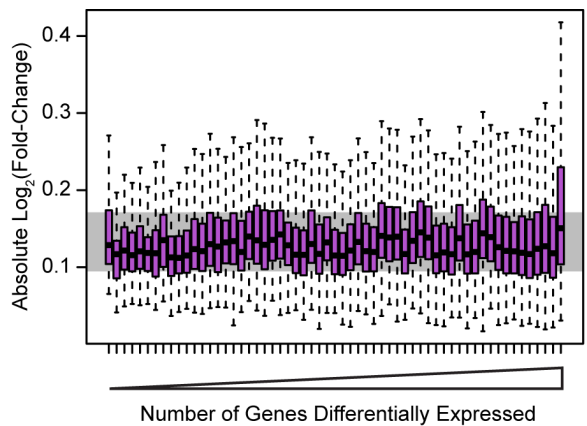

**Figure 2**

**Effect sizes for differentially expressed genes.** Boxplots of absolute $Log_2$(fold-change) between knockdown arrays and control arrays for all genes identified as differentially expressed in each experiment. Outliers are not plotted. The gray bar indicates the interquartile range across all genes differentially expressed in all knockdowns. Boxplots are ordered by the number of genes differentially expressed in each experiment. Outliers were not plotted.

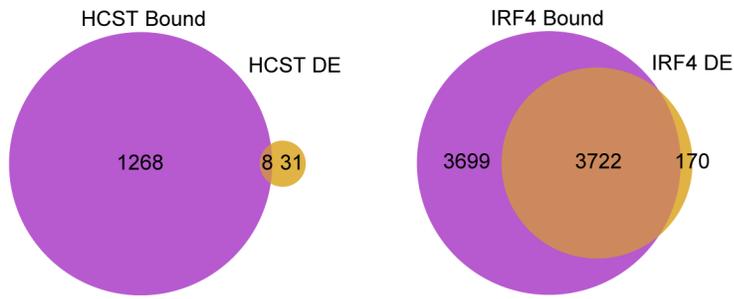
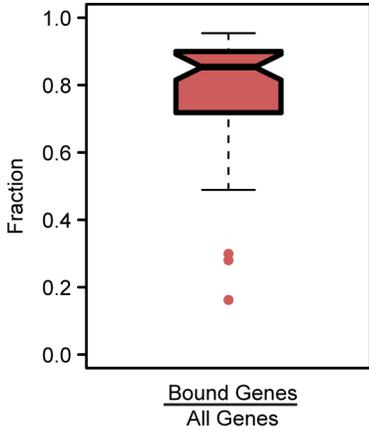
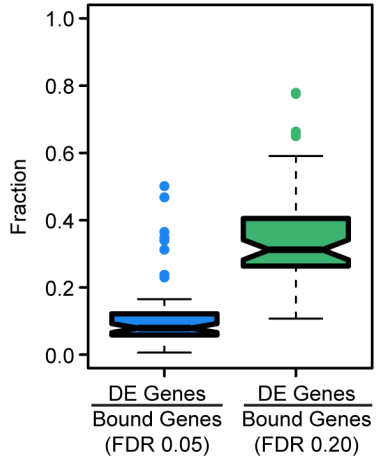

**Figure 3**

**Intersecting binding data and expression data for each knockdown.** (a) Example Venn diagrams showing the overlap of binding and differential expression for the knockdowns of *HCST* and *IRF4* (the same genes as in Figure 1). (b) Boxplot summarizing the distribution of the fraction of all expressed genes that are bound by the targeted gene or downstream factors. (c) Boxplot summarizing the distribution of the fraction of bound genes that are classified as differentially expressed, using an FDR of either 5% or 20%.

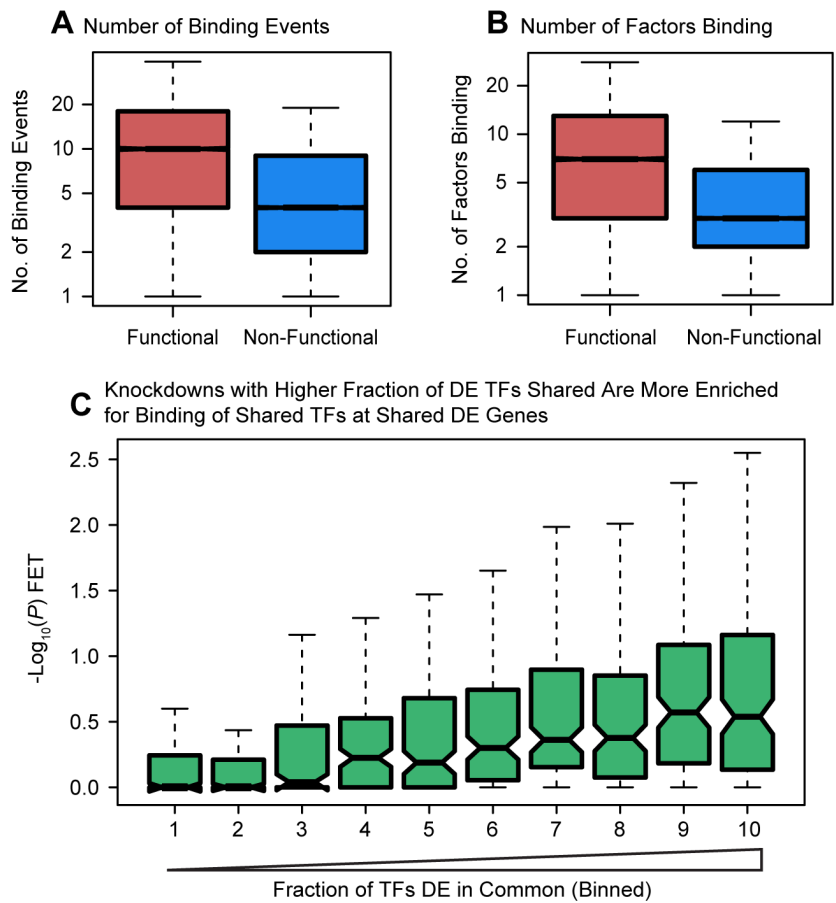

**Figure 4**

**Degree of binding correlated with function.** Boxplots comparing (a) the number of sites bound, and (b) the number of differentially expressed transcription factors binding events near functionally or non-functionally bound genes. We considered binding for siRNA-targeted factor and any factor differentially expressed in the knockdown. (c) Focusing only on genes differentially expressed in common between each pairwise set of knockdowns we tested for enrichments of functional binding (y-axis). Pairwise comparisons between knockdown experiments were binned by the fraction of differentially expressed transcription factors in common between the two experiments. For these boxplots, outliers were not plotted.

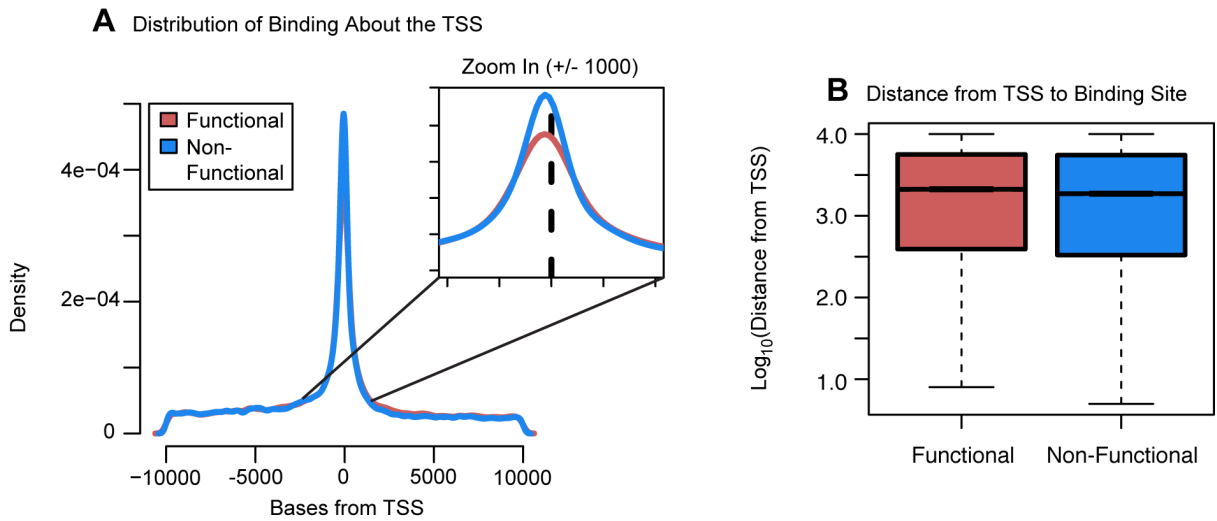

**Figure 5**

**Distribution of functional binding about the TSS.** (a) A density plot of the distribution of bound sites within 10kb of the TSS for both functional and non-functional genes. Inset is a zoom-in of the region +/-1kb from the TSS (b) Boxplots comparing the distances from the TSS to the binding sites for functionally bound genes and non-functionally bound genes. For the boxplots, outliers were not plotted.

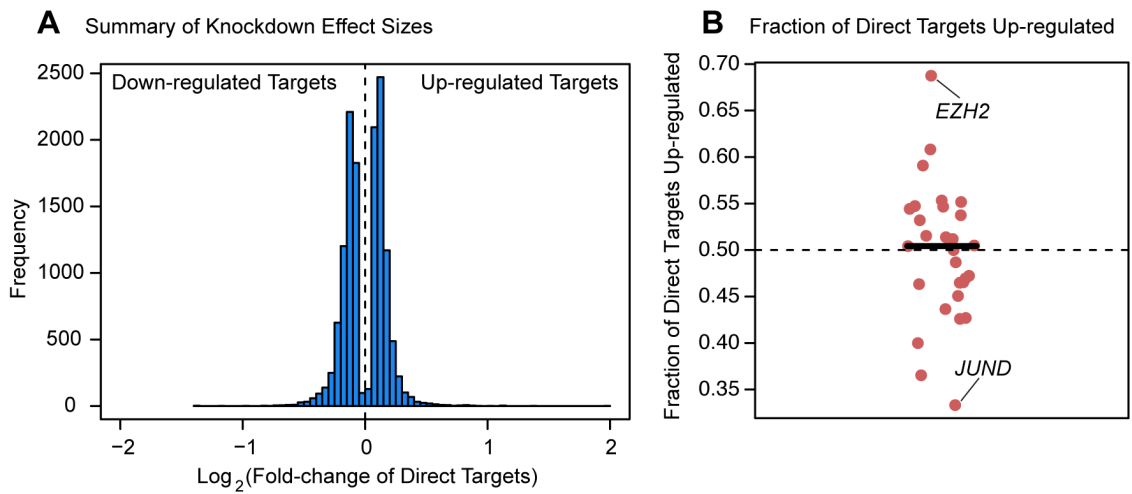

**Figure 6**

**Magnitude and direction of differential expression after knockdown.** (a) Density plot of all $\text{Log}_2$(fold-changes) between the knockdown arrays and controls for genes that are differentially expressed at 5% FDR in one of the knockdown experiments as well as bound by the targeted transcription factor. (b) Plot of the fraction of differentially expressed putative direct targets that were up-regulated in each of the knockdown experiments.